# The modular Atom Probe Concept

Patrick Stender[1,3], Helena Solodenko[1], Andreas Weigel[3], Irdi Balla[3], Tim Maximilian Schwarz[1], Jonas Ott[1], Manuel Roussell[1], Rüya Duran[1], Sebastian Eich[1], Mohammad Al-Shakran[2], Timo Jacob[2], Guido Schmitz[1,3]

1 Institute of Materials Science, Chair of Materials Physics, University of Stuttgart, Heisenbergstrasse 3, 70569, Stuttgart, Germany

2, Institute of Electrochemistry, Ulm University, Albert-Einstein-Allee 47, 89081 Ulm, Germany

3, Inspico, TTI GmbH, Nobelstraße 15, 70569 Stuttgart

Abstract

Atomic probe tomography (APT), based on the work of Erwin Müller, is able to generate three-dimensional chemical maps in atomic resolution. The required instruments for APT have evolved over the last 20 years from an experimental to an established method of materials analysis. Here, we describe the realization of a new instrument concept that allows the direct attachment of APT to a dual beam SEM microscope with the main achievement of fast and direct sample transfer. New operational modes are enabled regarding sample geometry, alignment of tips and microelectrode. The instrument is optimized to handle cryo-samples at all stages of preparation and storage. The instrument comes with its own software for evaluation and reconstruction. The performance in terms of mass resolution, aperture angle, and detection efficiency is demonstrated with a few application examples.

**Introduction.**

Nanotechnology is a key technology on which a variety of new products and improvements are based or will be enabled by. The ongoing trend towards technical miniaturization will require the control of chemical processes even on an atomic scale in the production of the smallest components in the future. There is already an urgent need for such high-resolution control in microelectronics, micromechanics and sensor technology, in the solar industry or in battery and fuel cell development, to name just a few examples. The targeted development of nanotechnological products and increasingly also quality assurance therefore require special microscopes that allow three-dimensional chemical analysis with atomic resolution. While the technological necessity is undisputed, the chemical local resolution capacity would be of no less value for biological systems and medical research. One method that can provide such information is atom probe tomography (APT).

The basic principle of APT is based on early work by Erwin Müller and his development of the field ion microscope (FIM) ((Müller 1951; Melmed 1996)). While the FIM provided the first atomically resolved images of metallic surfaces, the desire for a local chemical identification of the atoms drove the development forward for many years. By combining the FIM with a flight of time mass spectrometer, at least one-dimensional characterization of the sample became possible ((Müller, Panitz, and McLane 1968a; Panitz 1973; Müller, Panitz, and McLane 1968b; Panitz 1978)).

The step into three-dimensionality was achieved through innovative concepts and advances in computer, electronic and detector technology. In 1984, a three-dimensional volume reconstruction of the sample with atomic resolution was demonstrated for the first time at the University of Oxford (UK) by A. Cerezo and G. Smith ((Cerezo, Godfrey, and Smith 1988)) , and later (1992) with improved technology by B. Deconihout and D. Blavette ((Blavette et al. 1993)) at the University of Rouen (France). As a direct consequence, experimental systems were installed in various research facilities worldwide.

The first commercial 3D atom probes were offered by a spin-off company from the University of Oxford and under license from Cameca (manufacturer of physical measuring instruments). Taking a proposal from Nishikawa (Nishikawa et al. 1998) a new American company, Imago under Prof. T. Kelly (Kelly et al. 2004), began at the end of the 1990s to develop an innovative instrument that consistently used small extraction electrodes as an aid for evaporation of atoms and accelerated the measurement rate by two orders of magnitude. The last major development step was the usage of modern laser sources to support evaporation. This is based on early works by Kellog and Tsong (Kellogg and Tsong 1980) and was then re-recorded, first by the group in Rouen (Renaud et al. 2006) and shortly thereafter in the APT group in Münster from about 2004 (Schlesiger et al. 2010a; Oberdorfer et al. 2007; Gruber et al. 2009; P. Stender et al. 2007a). Since then, laser-assisted atom probes have been offered commercially by two competing companies IMAGO and CAMECA respectively. These were stand-alone highly developed instruments with ultra-high vacuum conditions. IMAGO was acquired by CAMECA (which in turn belongs to the American "Ametek" group). Since then, only one manufacturer of commercial atom probes is available. The LEAP instruments underwent a continuous evolutionary development, culminating in the previous state-of-the-art instrument of the Leap 5000 series. Many of these instruments have been installed in the world in recent years and have led to a surge in scientific working groups trying to use APT and to trace the laws that rule the atomic world.

Certainly, information gained from correlative microscopy, making use of the particular advantages of each methods to deliver complete information, enables scientists to derive reliable conclusions (Makineni et al. 2018). The possible sources of information are wide spread using X-rays, electron microscopes, scanning probe microscopy just to mention a few. But even during APT experiments, new sources of information are unlocked such as the microphotoluminescence (µPL) signals stimulated in semiconductors and insulating specimens, by the laser used intentionally for the laser enhanced field evaporation process (Houard et al. 2020). At the University of Rouen the photonic atom probe was developed, combining a laser assisted atom probe with an imaging spectrometer equipped with an CCD camera for time integrated analysis and an streak camera for time resolved analysis.

Running the atom probe as an independent, quite complex UHV system entails high investment costs. Also, sample preparation requires suitable tools and procedures. Samples need to be of acicular shape with an apex radius of about 30 to 100 nm. They can be produced by various preparation techniques. Electrochemical etching or sputter deposition of layer materials on pre-shaped tungsten wires (Patrick Stender et al. 2008) are more traditional approaches. The breakthrough, however, towards broad application of APT was made possible with the general use of FIB and the preparation methods developed with it (Prosa and Larson 2017). (Prosa and Larson 2017). Nowadays, the operation of an APT urgently requires the parallel operation of a suitable FIB instrument

Usually, focused ion beams are applied in a separate dual beam scanning microscope ("FIB"). The nanometric needles are then transferred from the FIB to the APT instrument, sometimes even between separate laboratories. This transfer is delicate and bears risks for unwanted side reactions or accident. Except just dropping of samples to the floor, these problems might be mechanical damage, surface oxidation, electrostatic charging, or unwanted phase changes.

Especially with sensitive materials or samples that need permanent cooling, the transfer from FIB to APT must be realized with suitable transfer systems (Stephenson et al. 2018).

In this article, we describe the layout, construction and development of an instrument that was designed to directly link the atom probe and the dual beam microscope. The synergy between both instruments allows new operational protocols and enables a quite economic solution, since the effort in UHV technology is significantly reduced. We benchmark the performance of this instrument by presenting exemplary measurements.

**Instrumental Design**

The concept of the modular instrument is based on our experience in construction and operation of conventional atom probe designs in the past (P. Stender et al. 2007a; Schlesiger et al. 2010b; Geber et al. 1992). The main goal of the concept was to shrink the APT to the absolute minimum and to gain synergetic advances that predominate possible disadvantages. In principle, this stripped APT just consist of a single vacuum chamber, a laser system, a high voltage supply and a suitable detector, all components that could be fitted as an additional equipment into an existing SEM or FIB system, if the vacuum levels would be compatible. But to improve usability and performance, a two-chamber solution was realized, since the time frame of an APT measurement is mostly incompatible to the normally short time slots in FIB usage. A functional core component is the dedicated transfer mechanism, the "APT shuttle" that enables exchanging samples in between the chambers in less than a minute. The design has been made compatible to most of the available focused ion beam instruments.

In the ongoing development, first a proof of concept demonstrator was constructed followed by a pre-series prototype. In this evolutionary process first improvements were already incorporated. Since the whole design is modular, additional equipment for dedicated sample treatments can easily be incorporated, also the design is open to adopt different detector or laser systems from a range of compatible technologies. The layout of the demonstrator and the pre-series model are shown in Fig. 1.

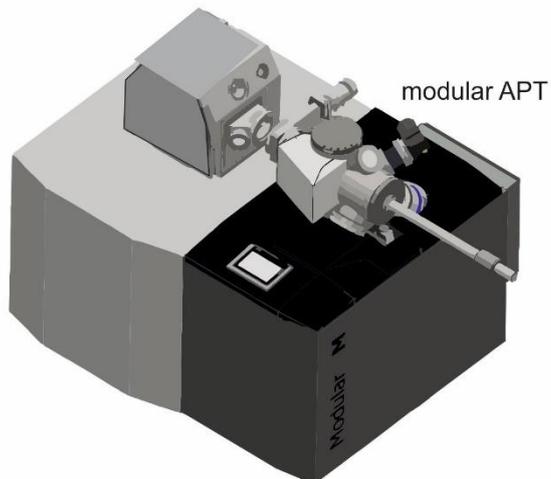
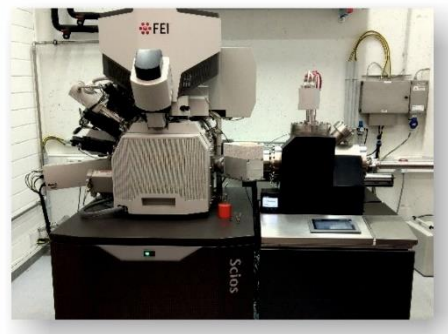
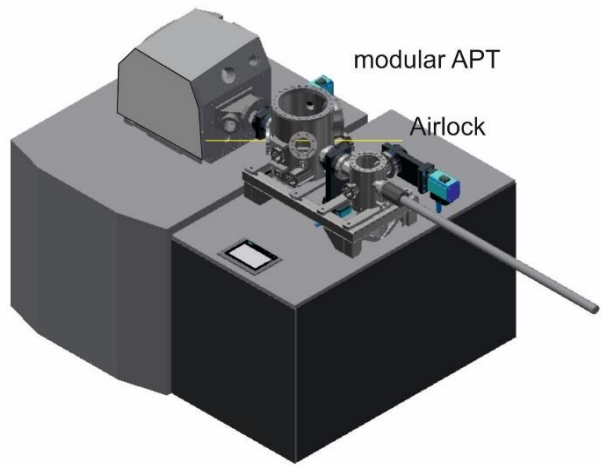
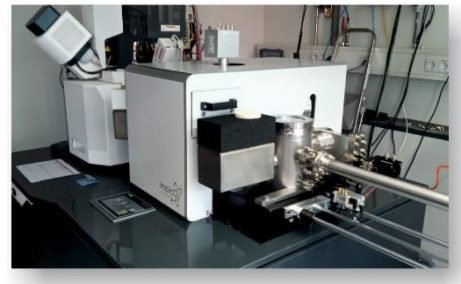

Figure 1 – a) Demonstrator Setup - Combination of a dual-beam microscope and a modular Atom Probe connected via the load lock port of the SEM chamber (installed at Univ. of Stuttgart, Germany). b) Second generation with airlock chamber and port to add a custom-designed experimental chamber (installed at University of Ulm, Germany). (Photographs with kind permission of INSPICO.)

**The APT Shuttle**

In the simplest case, a magnetic transfer rod would be sufficient to move a sample holder between the two chambers, if APT operation would follow conventional approaches. But we would like to make the instrument more flexible, so that the FIB can be also used for maintenance and alignment of the microelectrode. The central piece of the Modular-Atom-Probe has been miniaturized and computerized by attaching modern micromanipulators to a moveable microscope-head of 4 x 6 cm$^2$ footprint. This is the APT-Shuttle (Shuttle) (Fig. 2).

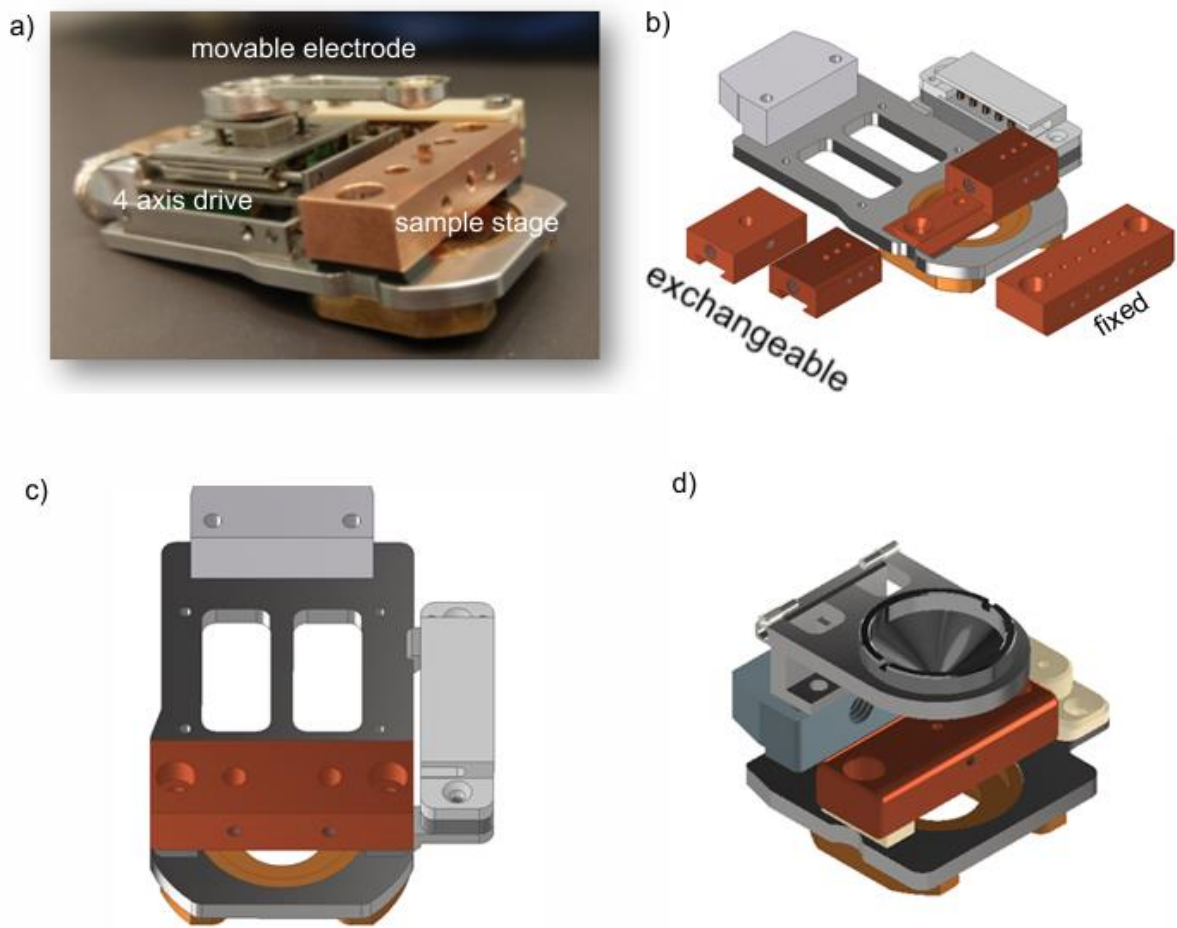

Figure 2 – a) The transfer sample stage including the *xyz* and ϕ movable extraction electrode, b) the sample stage can be modified and exchanged with regard to sample format (wires, TEm grids, cryo samples, SEM -stubs), c) base plate providing electrical connectors and setup room for different experiments, d) a robust version with mechanically adjusted macroscopic electrode version for student experiments.

This shuttle platform can be flexibly adopted to various experimental conditions as will be discussed later. The base plate provides an automatic electrical connector for high voltages up to 15 kV and additionally 10 more electrical contacts that can be used for multiple purposes. Currently they are used to provide the necessary connection to the micromanipulators to position the extraction electrode in space by four independent axes. (X, Y, Z and 360° Rotation). The stages inside the FIB and APT provide matching counter pin plates that ensure mechanical contact, when the shuttle is held in position by a dove tail system.

The sample supports themselves are easily exchangeable (Figure 2b) allowing adaptation to samples of various shapes. So far, single tips of metallic wire in copper, nickel or stainless-steel tubes as well as SEM stubs and TEM half grids are standard. Via automatic contact all shown supports are connected to the cryo-lines in the APT and SEM chamber as well. Two major categories of sample supports are offered. Fixed versions and interchangeable ones. The fixed ones allow mounting the samples and exchange the extraction electrode outside the instrument on a mounting bench. The interchangeable ones are equipped with a miniature dove tail underneath and can be inserted via an extra transfer line. This can be a cryo-suitcase or via an extra manipulator, allowing transfer to parking positions or reaction chambers. Examples for the closed cryo preparation routine will be described later.

In general, the shuttle can be modified in various aspects, if custom solutions for unique experiments are needed. For example, a robust mechanical version for the analysis of metal samples is depicted in Figure 3c. The mechanically adjusted macroscopic electrode provides a simple to use version without the danger to brake sensitive piezo drives by unexperienced users.

However, the addition of the piezo-driven extraction electrode to the shuttle offers the following possibilities:

1. The high-resolution capabilities of the SEM can be used to inspect the quality of the extraction electrode prior to measurement.
2. In the case of damage, the electrode can in some cases be machined by using the FIB beam for repair and to optimize measurement conditions.
3. The position of the extraction electrode can be adjusted using the high precision of the SEM. The tip position on the sample support is therefore flexible. Tips can be produced in-situ (Halpin et al. 2019) at various positions and the electrode is adjusted accordingly.
4. The mechanical coupling of tip and electrode enables a faster switching between SEM and APT, if intermediate SEM pictures of the tip are desired for calibration of the later reconstruction.

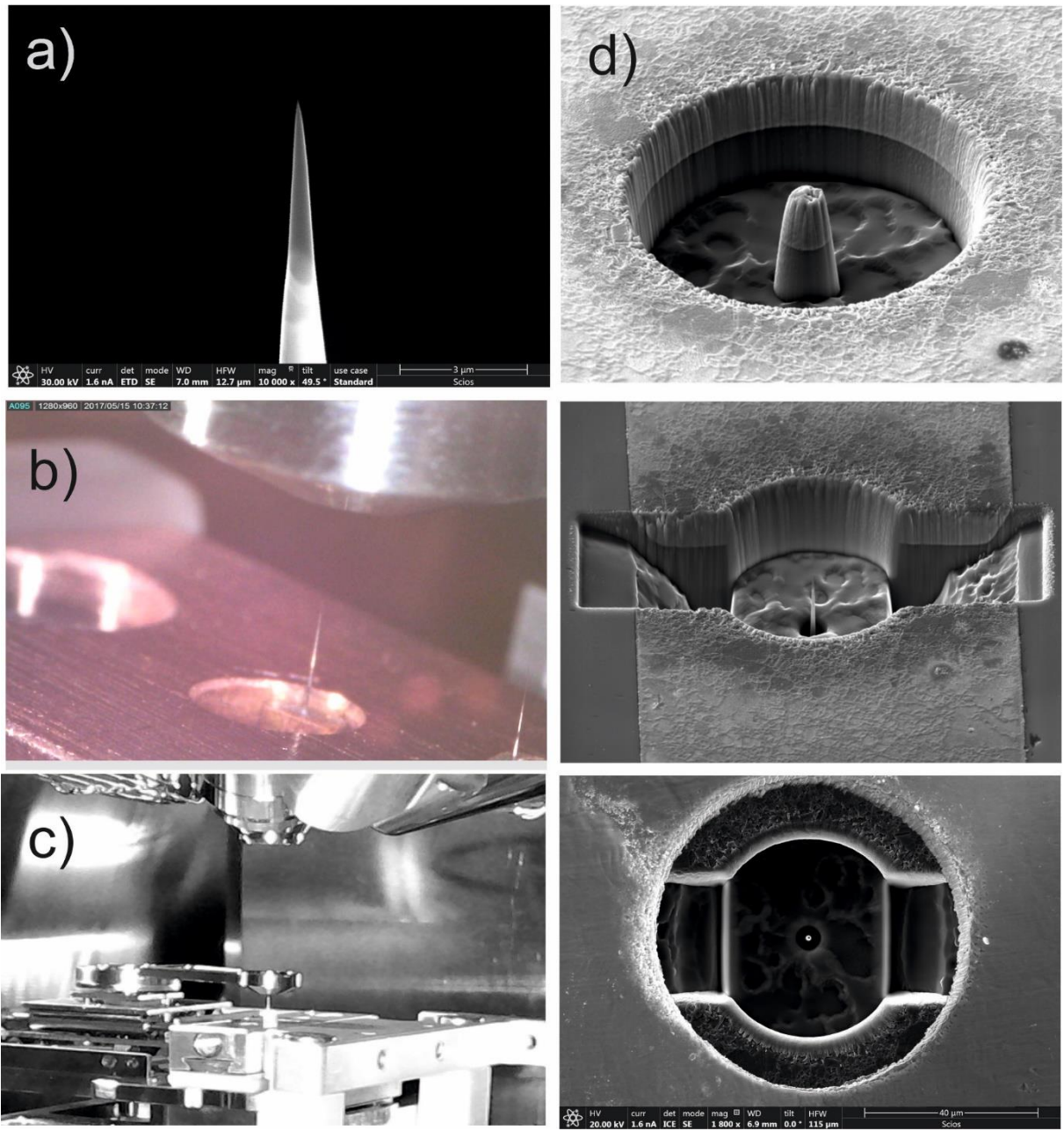

Figure 3- a) Prepared tip inspected by SEM; b) adjusted extraction electrode above sample viewed by APT chamber camera; c) adjusting the electrode inside the Dual-Beam instrument using the high resolution of the SEM; d,e) preparation of a tip without lift-out at arbitrary position and f) final electrode alignment (SEM view through the microelectrode)

**Instrument setup**

The APT part consist of single analysis chamber which is attached as a side chamber to the load lock of a commercial focused ion beam instrument. In order to combine both instruments a custom-made load lock adapter was developed bridging the gap between the Viton ring sealed vacuum system of the FIB and the UHV APT chamber. Since some of the commercially available cryo transfer systems are also installed onto the load lock port, the adapter has been designed to allow the addition of a Dewar for LN2 cooling (Fig 4 a) on the same port.

The APT measurement chamber uses an independent vacuum system. A pre-vacuum buffer is pumped by an oil free scroll pump in cycling operation while no APT measurement is running and in continuously operation while measurement mode is activated to create the necessary pre-vacuum for a turbo molecular pump. The residual gas pressure in the measurement chamber is in the $2*10^{-9}$ to $8*10^{-10}$ mbar pressure range, measured by ionization gauges. To decrease the vacuum level and to reduce hydrogen inside the chamber the system is equipped with a 500 L/s non-evaporable getter pump. This pump must be activated in dependence on the number of transfer processes, usually once a month. The trapped hydrogen is evaporated by heating the NEG pump for a certain time. All chambers are made from stainless steel based on the conventional CF flange sealing technique and can be baked up to a temperature of 120°C. The measurement chamber has a diameter of 200 mm and is designed for a vertical straight flight path. The horizontal axis allows transfer of the APT shuttle between APT and FIB. One laser entry and laser exit port are available. Two viewports CCD cameras allow monitoring the transfer and positioning of the samples inside the machine. Despite this, the necessary ports for the turbo-molecular-pump, non-evaporable-getter-pump, the magnetic transfer rod and feedthroughs for high voltage and cooling are included.

The APT shuttle is held in the measuring chamber via a dovetail system on a piezo-driven hexapod stage with 6 degrees of freedom (X,Y,Z, rotation x, rotation y, rotation z ) that can be adjusted independently to position and orient the tip with regard to the laser beam. The sample stage operates in a closed-loop mode. The smallest adjustment travel is 1 nm lateral and 1 urad angular resolution. The repeatability of the sample position is ±200 nm laterally after a complete traverse of the stage and the accuracy of the angular adjustment is ±10 urads. If the entire travel path is not used, the lateral repeatability is ±15 nm. Due to the high accuracy of the stage, sample positions can be approached very precisely. This allows the sample to be transferred back to the FIB to check the radius or other parameters. Afterwards the measuring position is found again quickly by addressing the stored coordinates of the hexapod. The *x* and *y* coordinates for the sample to be measured can also be automatically approached from the SEM coordinates of the samples by means of a coordinate transformation. The position within the atom probe is overdetermined by the 6 degrees of freedom. Therefore, all rotations are initially left at 0 and can be readjusted by the user if necessary.

The electrical contacts are made via special UHV compatible pressure spring contacts. The necessary cooling is achieved via an electro-mechanical unit that presses an electrically insulated cooling plunger, which is connected to the cooling head via copper strips, against the APT shuttle (Figure 4). Sufficient contact pressure is critical for the optimal cooling capacity of the actual tips. With the present design a minimum temperature of <30 K is obtained.

The initial design uses the FIB instrument as airlock and storage chamber, which reduces the complexity of the APT instrument. However, in a normal lab environment the FIB is used for multiple purposes. A dedicated FIB, which is just used for atom probe tomography, is seldom an affordable option. A normal atom probe measurement usually lasts several hours and in unique cases sometimes even several days. To avoid blocking the FIB from other service tasks during this time, an extra airlock chamber for the APT has been developed (Figure 4b), which is attached in extension of the transfer axis.

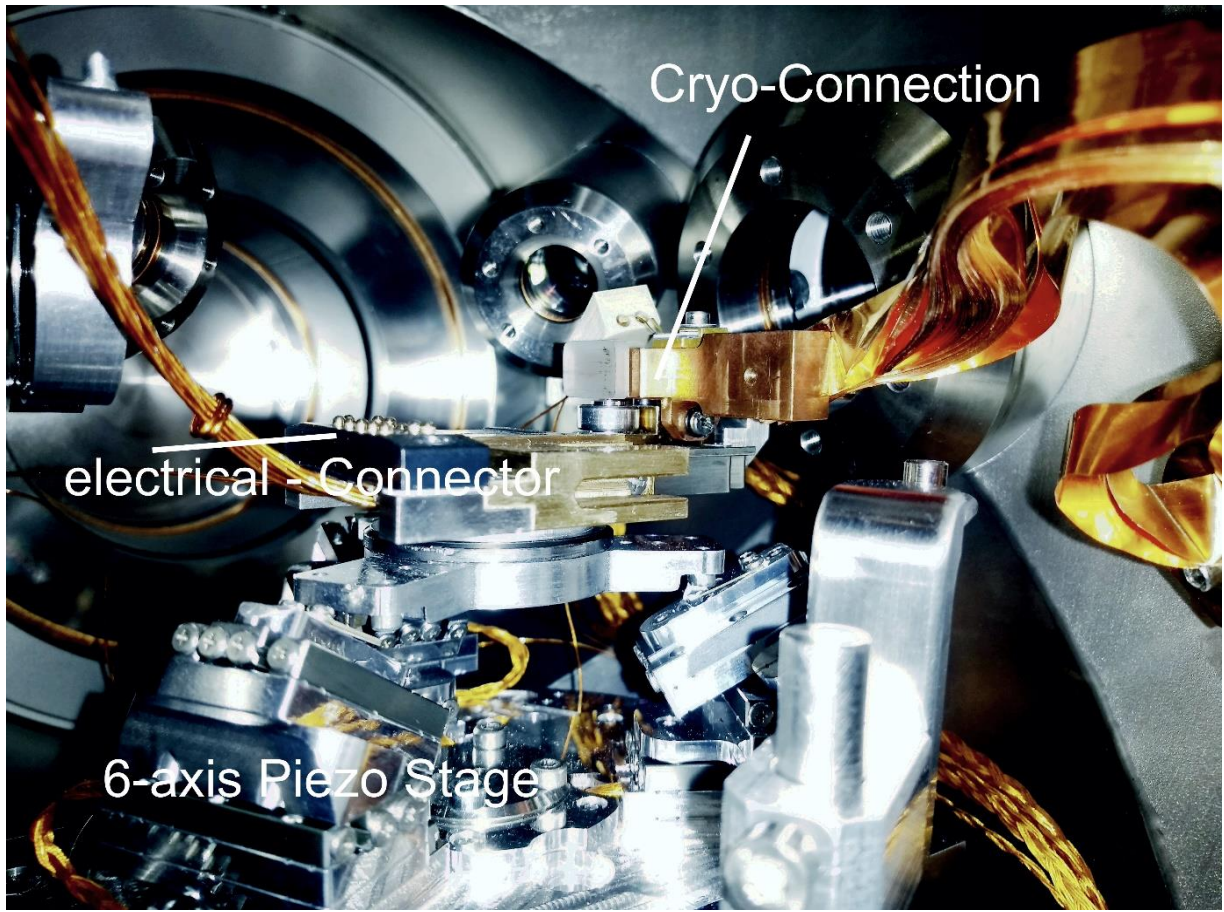

Figure 4– Inside view of the measurement chamber (Shuttle removed)

**The airlock chamber**

The airlock chamber has been designed for two purposes. The first purpose is to allow a fast transfer of samples into the APT chamber without interrupting ongoing experiments in the focused ion beam instrument. Although not necessary, it is convenient to have the option of inserting samples in multiple ways.

The chamber has been designed to hold multiple sample holders with a dove tail system on an internal stage. It thus also functions as a small storage unit. The special feature of the storage unit is that it can be connected to a Dewar via a flexible copper strip, which allows sample storage at about -140°C. (Note, the "airlock" is operated as a pure storage unit when using this cooling option.) For the remote replacement of the exchangeable sample holders on the shuttle, an axis is required which is arranged at a 90° angle to the main transfer axis. Also, this is operated via a small magnetic transfer rod (see Fig. 5 b).

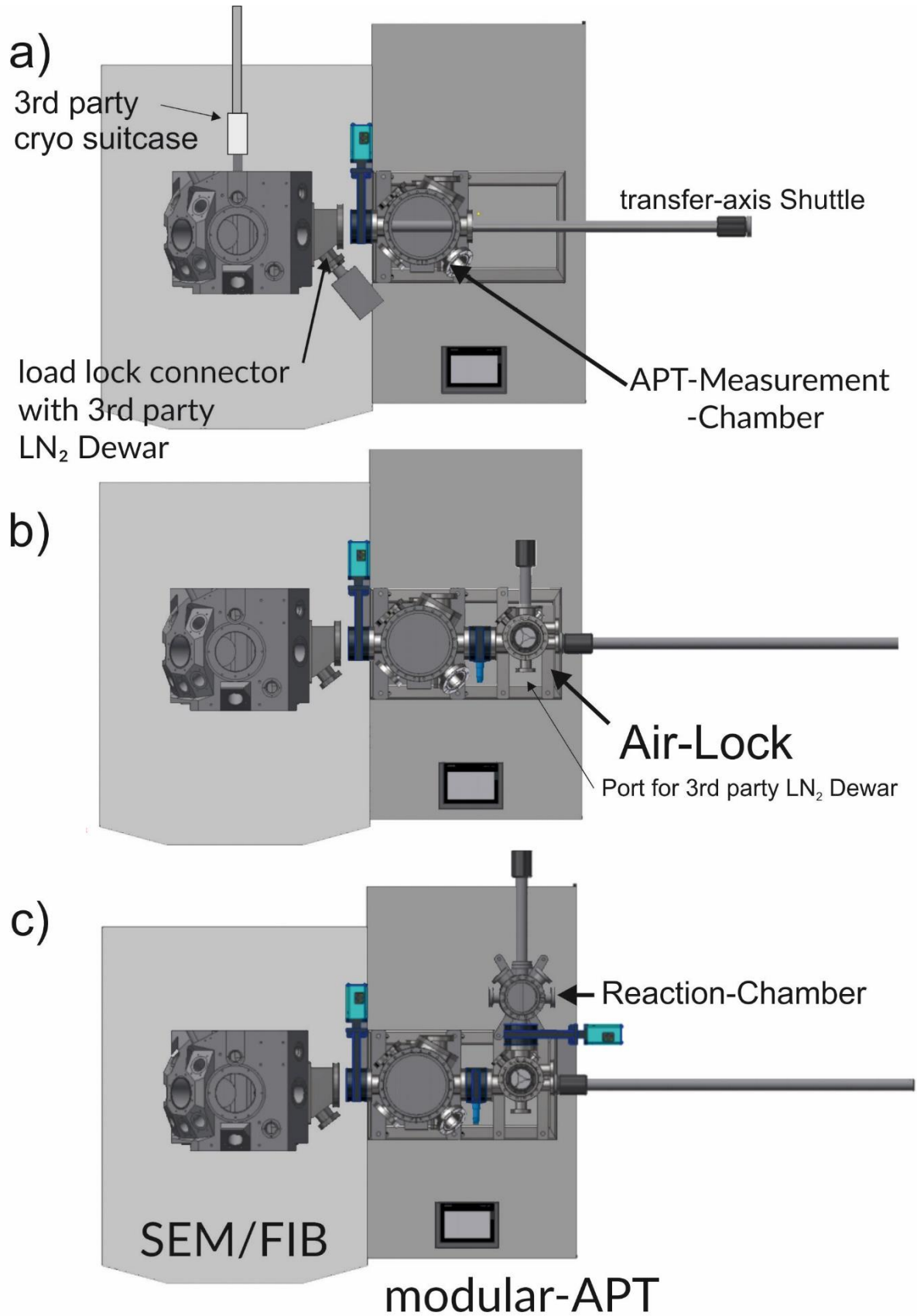

Figure 5: Different instrument setups: a) pure APT using FIB as airlock and preparation chamber, b) addition of a separate air lock chamber allowing independent loading of APT experiments, c) addition of a further side experimental chamber.

The second purpose of the air-lock chamber is to enable the addition of custom-made experimental chambers. Such experimental chamber can be located in extension of the previously mentioned sample holder transfer axis (see Fig 5c). Separated by a plate valve, an oven chamber, gas reaction or liquid corrosion cell or similar may be attached to the available port. The transfer rod is extended accordingly. Thus, an undisturbed transfer between FIB, measuring chamber, lock chamber and experiment chamber can be ensured without breaking the vacuum. Due to the existing cooled sample holders, this may even allow an uninterrupted cold chain of the samples, if the sample is also introduced cooled into the FIB by a suitable transfer unit. Of course, the air-lock chamber could also be upgraded to provide a load-lock for a VCT500 or similar commercial suitcase for direct transfer of cooled samples. The airlock chamber is equipped with a turbo-molecular pump connected to the common pre vacuum tank, reaching a base pressure of $10^{-8}$mbar.

**Cooling system**

The samples located on the shuttle can be cooled to a minimum temperature of 30K, when located inside the measurement chamber. A two stage Gifford-McMahon cryo head is used to achieve this temperature. Since vibrations are a very critical issue for the stability of the laser and even more for the resolution of the SEM, extra measures haven been undertaken to decouple and damp vibrations. The cryo head is mechanically decoupled using a bellow. Transmitted vibrations are further damped using polymeric vibrations damper in combination with oil pressure shock absorber. In addition, the entire chamber construction is mounted on air springs and decoupled from the FIB by another bellows and additional polymer dampers.

Nevertheless, reductions in SEM resolution are unavoidable when the cryo head is in operation. The resolution decreases to about 100 nm when the cryo had is in operation. Vibrational artefacts are detectable for slow scanning SEM pictures. Nevertheless, shaping of tips and TEM lamellas during operation of the APT cryo system is reliably possible. Possibly, further active damping measures could help to fully overcome this limitation.

Sample cooling inside the airlock chamber and inside the FIB is achieved via the usage of $LN_2$ Dewars connected by copper stripes to the respective sample stages. A cryo-stage has been developed for the FIB enabling the exchange of samples using a Leica sample suitcase (Leica VCT 500) and keeping a base temperature of -145°C. The suitcase has been modified to also accept the dove tail-based sample holder.

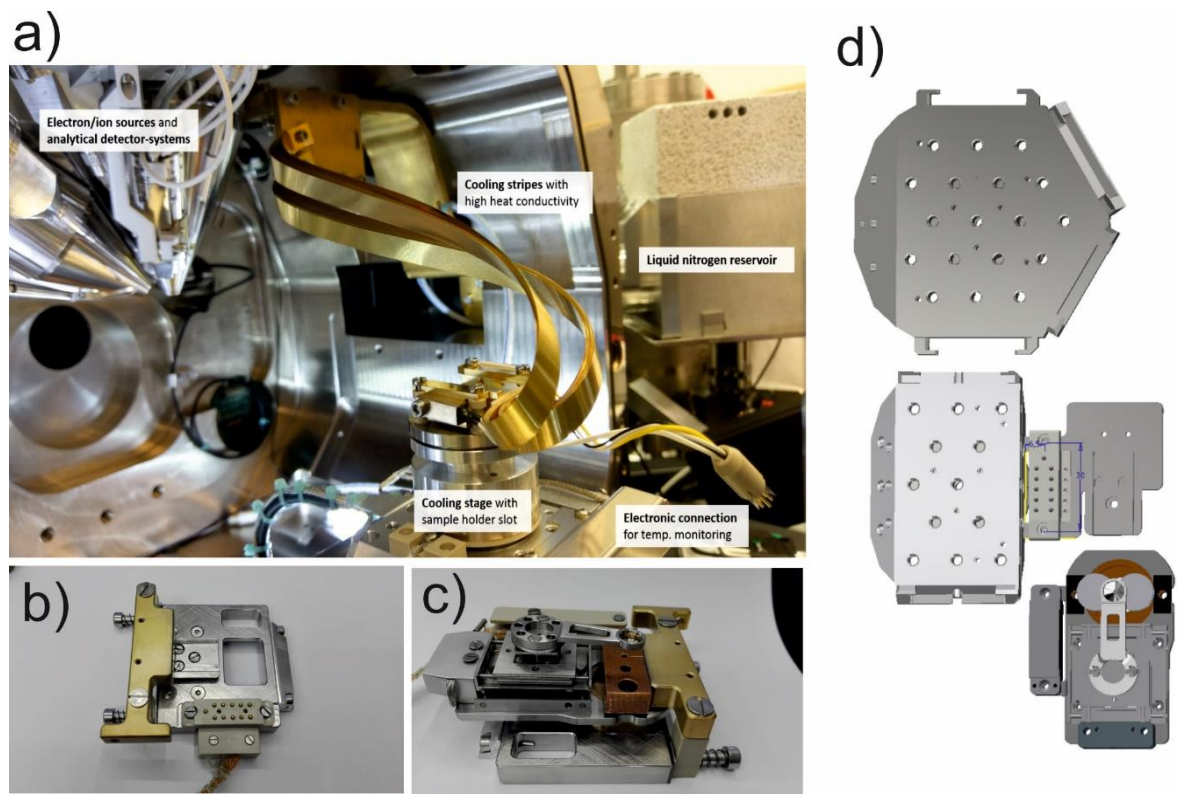

Figure 6: a) Leica Cooling setup for the FIB, dedicated cryo stage inside the FIB chamber without (b) and with (c) mounted APT shuttle under cryo conditions using the provided Leica cooling stripes and Dewar, d) (upper part) standard SCIOS sample stage for SEM stubs (lower part) modified sample stage to also accept the APT Shuttle without cryo option.

**Detector and HV system**

The detector system for the prototype instrument consists of a resistance matched chevron stacked MCP detector with an OAR ratio of >70% and an active diameter of >75 mm with a two-dimensional delay line. The MCP setup can be replaced by newly available funnel MCP with an open area ratio of 90% increasing ion detection efficiency above 80% (Fehre et al. 2018). Delay Line signals are amplified and processed by fast constant fraction discriminators with a maximum bandwidth of 300 Mhz and a double hit dead time of 10 ns. Digitized timing events are measured by a Time-to-Digital Converter (TDC) with 25 ps bin size resolution. For signals stemming from the MCP used as timing information for Time-of-Flight (TOF) spectroscopy, pulse intensity signals are recorded in addition. The system is prepared to also use a hexanode in view of high-voltage supplies, the chamber flange size and the control and filter algorithms.

**Laser system**

The laser pulses required for controlled evaporation are generated by industrial laser systems. The reliable laser sources are mounted below the chamber as a solid unit, together damped on air springs. While the demonstrator is equipped with a 10 ps pulse width laser, the second generation is equipped with a 350 fs laser system. Both systems use infrared as the fundamental wavelength (1034 nm) and provide both the second harmonic 517 nm and the third harmonic 344 nm. Switching between the wavelengths is possible for the experimenter by manually exchanging the polariser unit. Usually, however, the third harmonic oscillation is used for the experiments. The sample is focused to the laser spot in the chamber via UV sensitive camera optics arranged similar to a confocal microscope. The final lens is located inside the vacuum chamber with a short focal length of 40 mm. The spot size achieved is currently < 8 μm in dependence of the chosen beam expansion factor. The laser spot can be scanned in a limited area via a high-frequency piezo tilting mirror system. The scanning system has a resolution of 0.2 µrad and a repeatability of 0.4 µrad and operates in close-loop mode with a response time in the ms range.

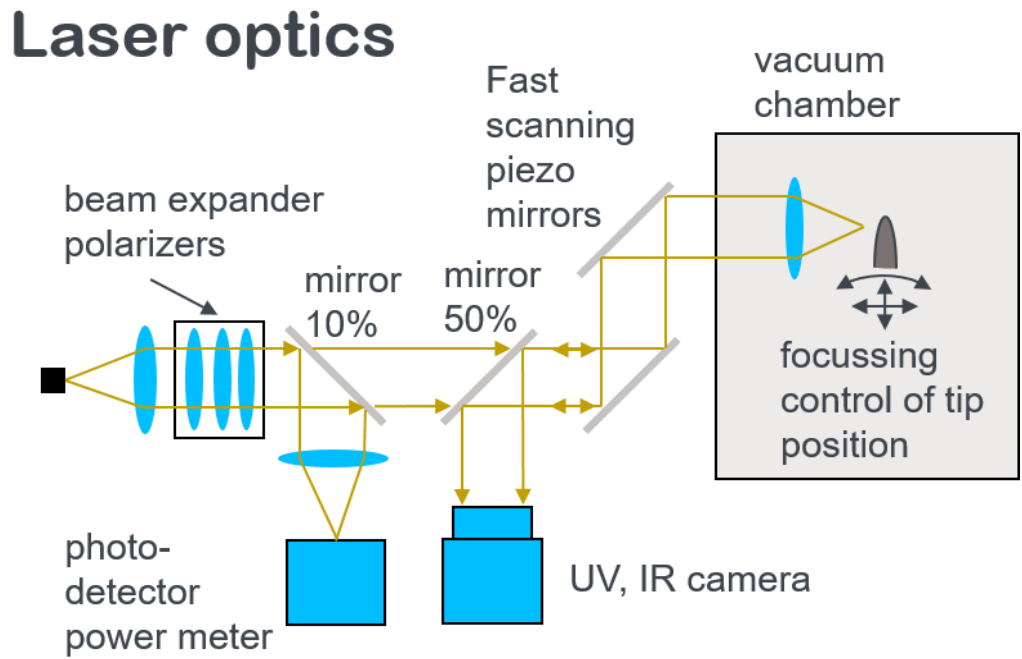

Fig. 7: Laser system

**Characteristics**

With a flight length of 105 mm, the geometrical opening angle accounts to ± 20°. Due to image compression arising from the tip geometry, the observable opening angle is larger, as can be derived from the pole figure observable from a tungsten measurement depicted in Figure 8. The poles (011) and (112) are resolved. Using them for calibration, the image compression factor $\kappa$ is determined to 0.54 in accordance to earlier determined values (P. Stender et al. 2007b) (P. Stender et al. 2007b), which fixes the effective aperture angle to ± 37°.

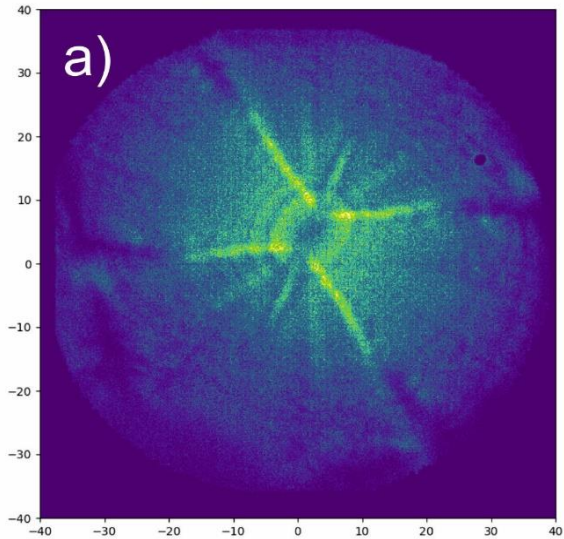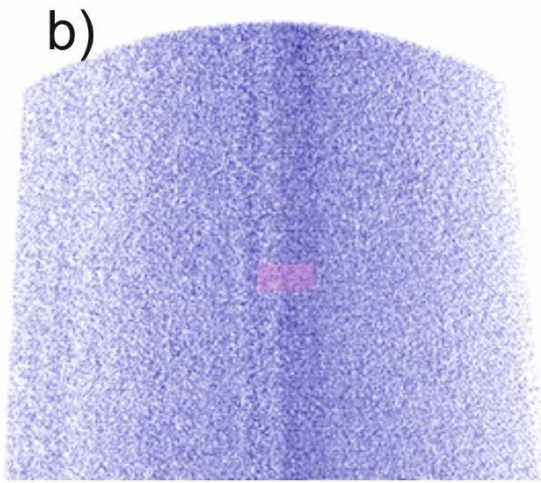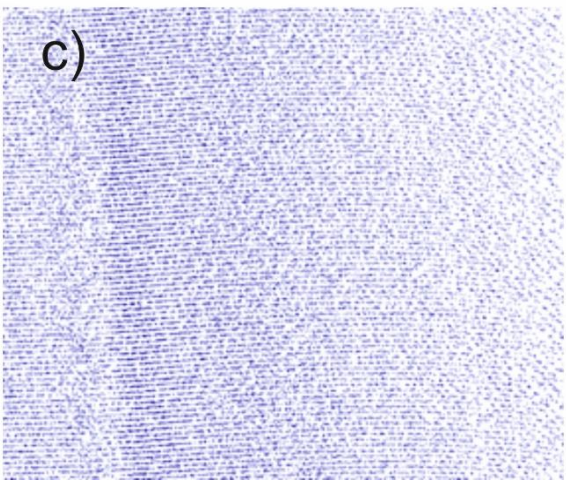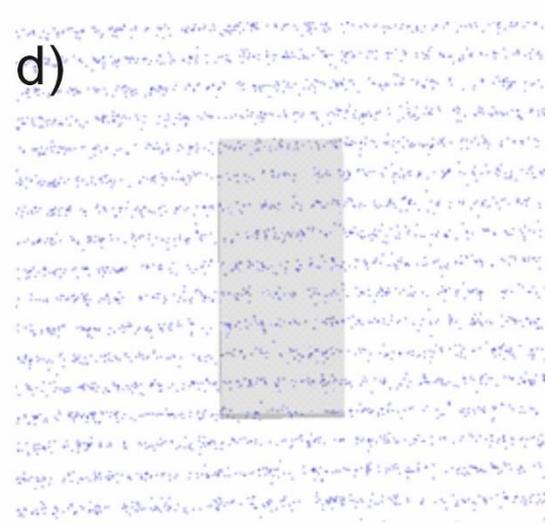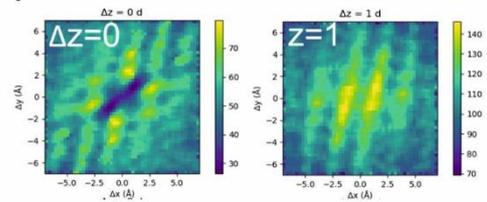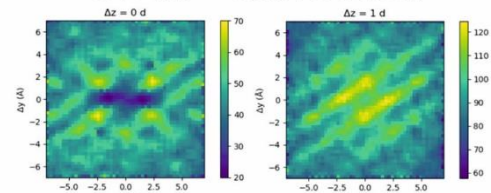

Figure 8: a) Desorption map of tungsten, b) reconstruction of a tungsten measurement revealing lattice planes, c) magnified view demonstrating two sets of lattice planes , d) lattice planes zoomed in

(inset dimension 1 nm x 2nm), e) spatial distribution maps calculated in x-y plane, f) calculated lattice spacing along z-axis

The mass resolution was tested by measurements on tungsten, aluminium and nickel samples, among others. The mass resolution m/Δm, determined over the full width of the detector, is for the aluminium measurements with FWHM of (0.039 ± 0.001) u at m = 27 u about 700 (692±20). Somewhat lower values were found for nickel with m/Δm = 612 ± 12 and tungsten with m/Δm = 683 ± 10. The signal-to-noise behaviour spans a dynamic range of up to 3.5 orders of magnitude.

The open area ratio (OAR) of the used microchannel plate is specified to be 70%. In order to determine the detection efficiency, the reconstructed volume must correctly match the lattice properties of the real physical sample. While the z-resolution is typically sufficiently high to resolve the individual lattice planes and, thus, the required z-scaling, the correct scaling of the lateral directions is less obvious. It was shown by Geiser et al. maps (Geiser et al. 2007) that spatial information in the lateral directions can still be obtained from atom probe data using so-called spatial distribution maps (SDMs).

An SDM is a two-dimensional histogram visualization of all interatomic distance vectors

$$\Delta \vec{r_{ij}} = (\Delta x_{ij} \Delta y_{ij} \Delta z_{ij}) \quad (1)$$

where the lateral components, i.e. $\Delta x_{ij}$ and $\Delta y_{ij}$ are chosen here for the two-dimensional histogram data. Depending on the lattice type and orientation, preferred combinations $\Delta x_{ij}$, $\Delta y_{ij}$ exist which will be seen as maxima in the histogram.

As shown by Geiser et al. (Geiser et al. 2007), the lateral resolution is optimised if the $\Delta x_{ij}$, $\Delta y_{ij}$ are chosen for atomic pairs within a very narrow $\Delta z$ -separation. In a first step, the sample volume is optimally aligned in z-direction using statistical criteria, from which also the z-scaling factor for the correct lattice spacing is obtained. Then all pairs $\Delta x_{ij}$, $\Delta y_{ij}$ are chosen for which

| | |
|---|---|
| $\Delta z_{ij} < 0.1\ Angström$ | 2 |

holds. The resulting histogram is shown in Fig. 8e upper part after application of a Wiener filter for data smoothing.

As can be seen, lateral information does exist in this histogram, however, the lattice is obviously distorted. Since the real crystallographic positions of these maxima are known, the histogram can be transformed by taking two linearly independent maxima (determined by two-dimensional Gaussian fits) and solving a system of linear equations to obtain the transformation matrix. The result is shown in Fig. 8e lower part. With knowledge of the real volume corrected this way, the detection efficiency is obtained as 68% ± 8. (Note that lateral information is also visible for atomic pairs which do no stem from the same lattice plane $\Delta z = 0$, but also for pairs from neighbouring planes with

| | |
|---|---|
| $\Delta z = d_z$ | 3 |

where $d_z$ is the lattice spacing in z-direction. The determined detection efficiency agrees with the value expected from the OAR of the MCP within the limits of error accuracy. Similar results have been reported earlier (Fehre et al. 2018).

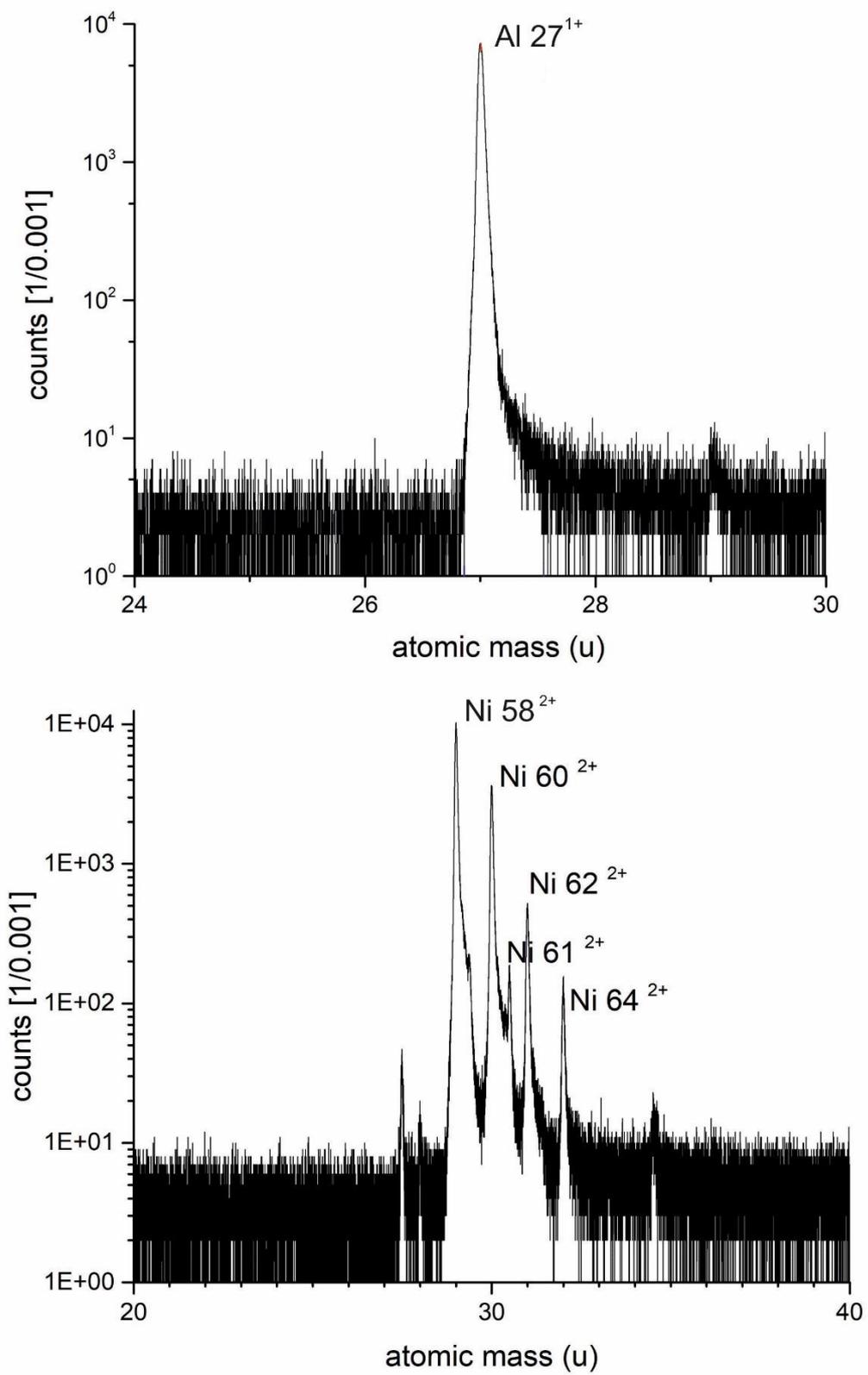

Figure 9: Mass spectra of an exemplary Aluminium (upper spectra) and a Nickel sample.

**Experimental**

The samples are prepared and the microelectrode is aligned in the FIB before measurement. The shuttle is then transferred to the atom probe chamber. An experienced operator needs about 30 s for the complete mechanical transfer. However, the different vacuum levels prove to be critical here. Ordinary FIB instruments work in a pressure range of $10^{-4}$ to $10^{-6}$ mbar. An improvement of the vacuum into the high $10^{-7}$ mbar range is observed when operated in cryo experiments. The typical working pressure in the atom probe is between $10^{-9}$ and $10^{-10}$ mbar. For the transfer of the shuttle, the intermediate plate valve must be opened, so that there is an inevitable deterioration of the vacuum level in the APT due to the transfer process. A regeneration time of about 30 minutes is observed to return to the medium $10^{-9}$ mbar. However, the risk of thawing for temperature-sensitive samples is avoided by the permanent cooling inside the APT. Even if the actual transfer takes only a few seconds, an immediate start of high-quality measurements after the transfer has to be avoided due to the vacuum restrictions.

On the good side, the instrument offers new opportunities for a quick sample check if the mass spectrum suddenly shows a deterioration during the measurement. As an illustrative example, Figure 10a shows an aluminium tip before the measurement. In the SEM picture the typical vibration artefacts are seen that are caused by the cooling head. The sample was analysed for about 10 million measured atoms (see reconstruction in inset of Fig. 10a), when the signal to noise ratio collapsed drastically. A flashover or a break-off of the sample was expected. To check, the sample was transferred back to the FIB for inspection (Figure 10 b) showing directly the damage of the sample probably caused by an electrical flash. Since there was still enough material left, the sample was then directly sharpened again and the measurement was continued after about 30 minutes total processing time (see Figure 10 c) to count another $10^7$ atoms. For even faster transfer and measurement, a dedicated UHV FIB would be a viable option.

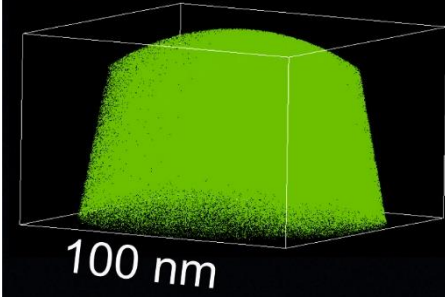
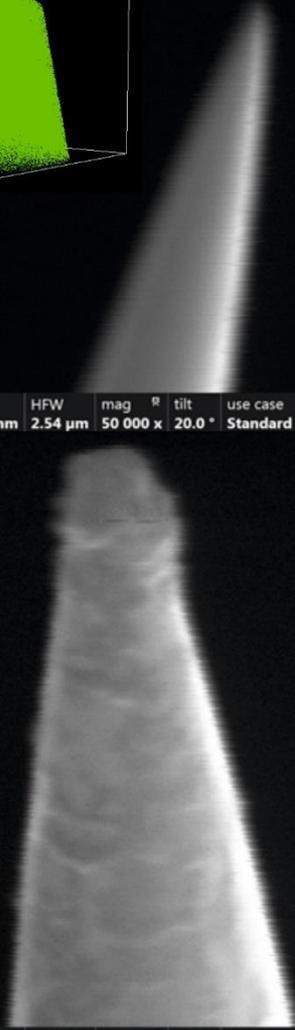
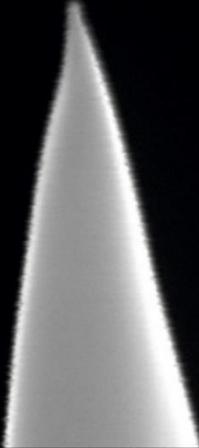
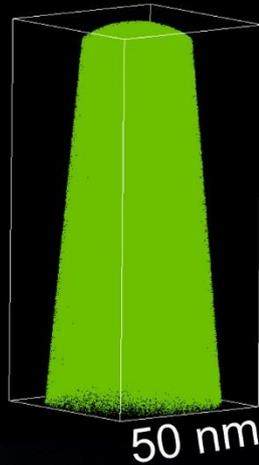

Figure 10: a) Aluminum sample before measurement, b) after flashing event, c) after reshaping.

**Reconstruction-Software**

For the reconstruction of the data, an independent software solution based on OpenGl and C++ was developed (Stender, P., Balla, n.d.). The raw data are stored in a binary data format that was also used by the previous machines (P. Stender and Schmitz 2006; Schlesiger et al. 2010a). Other file formats such as Pos and Epos are supported ((Gault et al. n.d.). However, only the raw format allows the full use of all functions so far. The software is being continuously developed to be opened for other data formats as proposed by the IFES technical committee. The software offers simple filtering options of the raw data, mass spectra with bowl correction, a graphical alloy editor with SQL database for already defined molecules. The tips can be reconstructed with different reconstruction modules following the protocols of (Bas et al. 1995) and (Jeske and Schmitz 2001)). For the geometric reconstruction algorithms there is the additional possibility to use parameterised curves for the shaft angle of the tip. For the filters as well as for the reconstruction parts of the software there is the possibility to insert own modules in Java-Script and to manipulate the data accordingly. Various export formats allow the data to be transferred to other software solutions. After successful reconstruction, the peak is displayed as a colour-coded point cloud as usual. The representation of the data can be adjusted by manipulating the data stream using element and geometry filters. Data evaluation currently includes iso-surfaces and orthoslices, composition profiles, cluster search algorithms, nearest neighbour analyses and spatial distribution maps (Geiser et al. 2007). The aim is to further expand the number of possible evaluation modules.

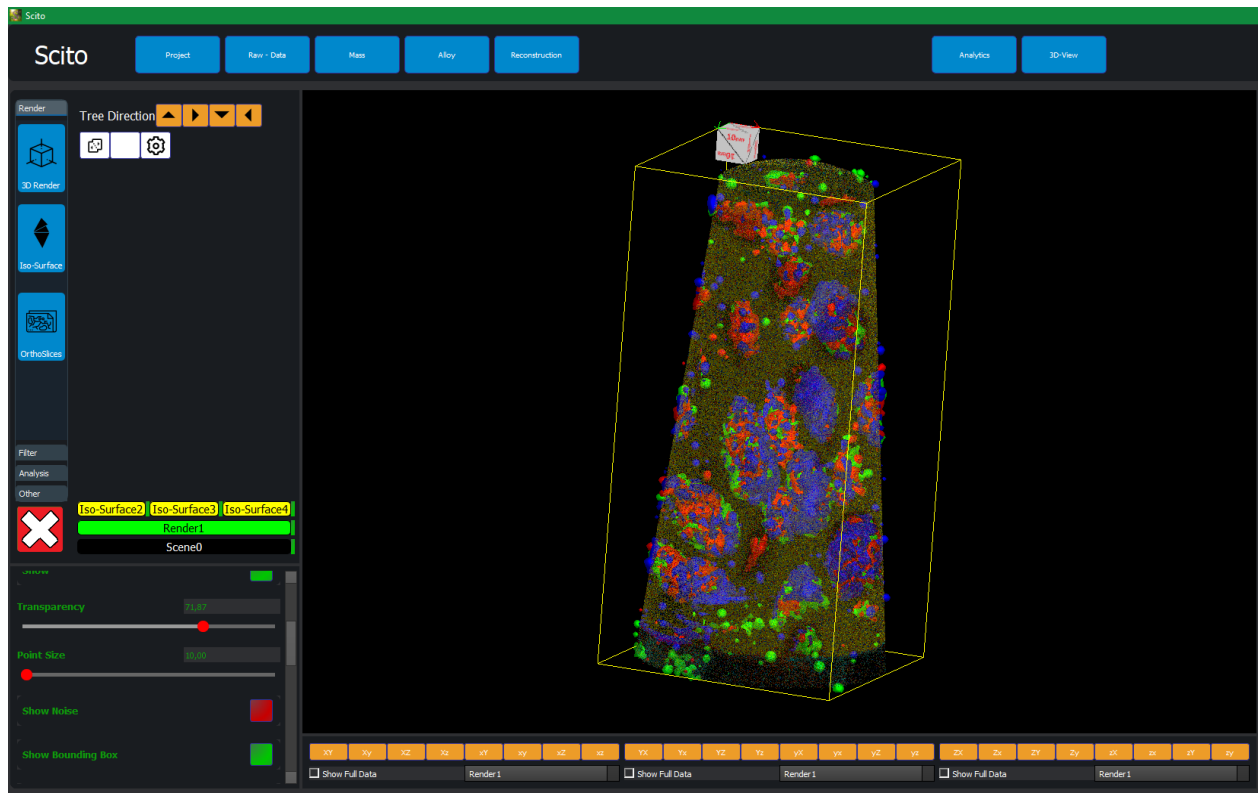

Figure 11 – Reconstruction software Scito. The 3D reconstructed tip is shown here. Iso-surfaces of different colour reveal Ti (green), Al (red) and Ni(blue) localization in the precipitation process

**Summary**

In established atom probe analysis, the APT for controlled field desorption and the FIB microscope for sample preparation are operated as two separated instruments. In order to raise economic advantages of synergy, to enable fast cryo-transfer and to introduce new operational modes in maintenance and alignment of extraction electrodes and tip, we presented a new instrument that represents basically a direct combination of APT and dual-beam scanning microscope. The necessary internal and mechanics to realize a fast cryogenic sample transfer has been developed and demonstrated. Sensitive samples can be prepared and transferred to the APT without breaking the vacuum and without an additional transport device within a few seconds. The same applies to frozen liquids or samples that need to be permanently cooled and protected from atmospheric influences.

Due to the fast transfer between the units, flashed samples can be easily re-tipped and further measured. SEM images before and after the measurement can conveniently provide information about the actual evaporated volume. The extraction electrode, which is often affected by dirt or electrical flashovers, can be inspected easily in the scanning electron microscope and, if necessary, reworked by the ion beam.

The atom probe performance data in terms of mass resolution, aperture angle, detection efficiency and signal to noise ratio demonstrate a reliable analysis without compromises in comparison to dedicated stand-alone instruments.

Remaining mechanical vibrations transferred from the cryo-stage to the FIB and long vacuum restoring time after transfer are presently the main limitations, but could be certainly further improved by raising the vacuum quality in the FIB and decoupling the cryo head via a gas reservoir.